# Probing computational methodologies in predicting mid-infrared spectra for large polycyclic aromatic hydrocarbons

B. Kerkeni,[1,2,3]★ I. García-Bernete,[1]★ D. Rigopoulou,[1]★ D. P. Tew,[4] P. F. Roche[1] and D. C. Clary[4]

[1]*Department of Physics, University of Oxford, Oxford OX1 3RH, UK*
[2]*ISAMM, Université de la Manouba, La Manouba 2010, Tunisia*
[3]*Département de Physique, LPMC, Faculté des Sciences de Tunis, Université de Tunis el Manar, Tunis 2092, Tunisia*
[4]*Department of Chemistry, University of Oxford, Oxford OX1 3QZ, UK*

## ABSTRACT

We extend the prediction of vibrational spectra to large sized polycyclic aromatic hydrocarbon (PAH) molecules comprising up to ~1500 carbon atoms by evaluating the efficiency of several computational chemistry methodologies. We employ classical mechanics methods (Amber and Gaff) with improved atomic point charges, semi-empirical (PM3, and density functional tight binding), and density functional theory (B3LYP) and conduct global optimizations and frequency calculations in order to investigate the impact of PAH size on the vibrational band positions. We primarily focus on the following mid-infrared emission bands 3.3, 6.2, 7.7, 8.6, 11.3, 12.7, and 17.0 μm. We developed a general Frequency Scaling Function ($\mathcal{FSF}$) to shift the bands and to provide a systematic comparison versus the three methods for each PAH. We first validate this procedure on IR scaled spectra from the NASA Ames PAH Database, and extend it to new large PAHs. We show that when the $\mathcal{FSF}$ is applied to the Amber and Gaff IR spectra, an agreement between the normal mode peak positions with those inferred from the B3LYP/4-31G model chemistry is achieved. As calculations become time intensive for large sized molecules $N_c > 450$, this proposed methodology has advantages. The $\mathcal{FSF}$ has enabled extending the investigations to large PAHs where we clearly see the emergence of the 17.0 μm feature, and the weakening of the 3.3 μm one. We finally investigate the trends in the 3.3 μm/17.0 μm PAH band ratio as a function of PAH size and its response following the exposure to fields of varying radiation intensities.

## 1 INTRODUCTION

The investigation of polycyclic aromatic hydrocarbons (PAHs) properties is relevant in several different research fields. In particular, they have been extensively invoked in combustion chemistry (Richter & Howard [2000]) as they play an important role in atmospheric pollution and present toxic functions. In astronomy, PAHs are detected (Tielens [2011]) in a variety of space environs via their spectral fingerprints. Observations with the Infrared Space Observatory (*ISO*) and the Spitzer have shown since the mid-90s, among them fullerenes have also been directly identified (Herbst [2006]). PAHs can be considered as finite graphene sheets passivated with hydrogen atoms to form either catacondensed PAHs with all of their carbon atoms situated in one or a maximum of two molecular constitutive rings, or pericondensed PAHs with some of the carbon atoms situated in more than two molecular constitutive rings. They play an important role and are an ubiquitous component of organic matter in space. Their contribution is invoked in a broad spectrum of astronomical observations that range from the ultraviolet to the far-infrared and cover a wide variety of astronomical objects ranging from meteorites and other Solar System bodies to the diffuse interstellar medium, in the local Milky Way and in external galaxies. Strong emission features dominate the mid-infrared (MIR) spectra of our own Galaxy, galaxies in the local Universe as well as distant galaxies exhibiting an overall similar spectral profile among different sources (Li [2020]). Distinct emission peaks at 3.3, 6.2, 7.7, 8.6, 11.3, 12.7, and 17.0 μm appear with weaker and blended features distributed in the 3–20 μm region. The 6–9 μm region, in particular, contains three well-known major features: a band at 6.2 μm, a large complex of multiple overlapping bands at approximately 7.7 μm, and a band at 8.6 μm. Decomposition of the 7.7 μm complex using either Gaussians or Lorentzians identified several sub-components with the dominant ones centred around 7.6 and 7.8 μm (Bregman et al. [1989]; Cohen et al. [1989]; Verstraete et al. [2001]; Kerckhoven [2002]; Peeters et al. [2002]). These features are generally attributed to infrared (IR) fluorescence of PAH molecules pumped by UV photons followed by internal conversion and emission through a fluorescence cascade.

IR spectra of PAHs have been the subject of several observational, experimental and modelling studies over the last decades e.g. (Van Dishoeck [2004]; Herbst [2006]; Salama [2008]; Tielens [2008]; Cami et al. [2010]; Candian & Sarre [2015]; Buragohain et al. [2020]; Li [2020]; and references therein), in order to identify relevant molecular structures and characterize their physical and chemical conditions in various astrophysical environments such as interstellar and circumstellar, galactic and extra-galactic, and in the early Universe.

★ E-mail: boutheina.kerkeni@physics.ox.ac.uk (BK); ismael.garciabernete @physics.ox.ac.uk (IGB); dimitra.rigopoulou@physics.ox.ac.uk (DR)

To date, the interpretation of several observed PAH features is still poorly understood, and the carriers are tentatively assigned either to hydrogenated PAHs (Schutte, Tielens & Allamandola 1992; Bernstein, Sandford & Allamandola 1996), or to PAHs with aliphatic side groups (Kondo et al. 2012; Buragohain et al. 2020), nitrogenated PAHs (Bauschlicher et al. 2018), and with metal complexes (Simon & Joblin 2009). Other authors have considered side groups attached to PAHs such as polyphenyls (Talbi & Chandler 2012), vinyl groups, Boron-nitrogen coronene and boron-phosphorous coronene (Maurya et al. 2012). Other theoretical modelling of vibrational and electronic absorption spectra (de Abreu & Lopez-Castillo 2012) were explored with halogen atoms like fluorine, chlorine, bromine and iodine (Gardner & Wright 2011), and protonated PAHs (Bahou, Wu & Lee 2012). Triplet spin states in the case of dehydrogenation states of PAHs were investigated experimentally and computationally (Pauzat & Ellinger 2001; Galué & Oomens 2012).

In some astrophysical objects PAHs are believed to account for about 20 per cent of carbon, and can grow to immense sizes including hundreds of atoms (Allain, Leach & Sedlmayr 1996; Tielens 2005; Allamandola 2011), due to their resistance to strong UV flux in nearby star-forming regions. Presently, the detection (Cami et al. 2010) of $C_{60}$ motivated us towards performing computations for large molecules. These findings motivate researchers to elucidate the underlying formation mechanisms as both large aromatic systems and a small aromatic molecule have been identified by astronomical observations in combination with laboratory spectroscopy (Campbell et al. 2015; Cernicharo et al. 2001; McGuire et al. 2018). In their study of the two reflection nebulae NGC 7023 and $\rho$ Ophiucus-SR3, (e.g. Rapacioli, Joblin & Boissel 2005) showed that PAH molecules are produced by the decomposition of small carbonaceous grains inside molecular clouds, these grains being interpreted as PAH clusters. A minimal size of 400 carbon atoms per cluster was inferred from the analysis of astronomical observations (Tielens 2008). We therefore extended the computational investigations to PAHs larger than what is currently available in the literature. Employing a pool of Quantum Chemical (QC) methods, we provide an inventory of MIR spectra relevant from small to large PAHs, which are believed to be ubiquitous in the spectra of galaxies near and far. We are mainly interested in the investigation of band strengths ratios.

As they exhibit delocalization of the $\pi$ electrons and polarization effects related to their extended electronic conjugation coupled with their planar structures, PAHs are challenging systems for density functional theory (DFT), particularly when their size increases. Some authors (Savarese et al. 2020) carried out investigations of spin density in seven open-shell PAH systems by testing a range of different density functional approximations, they showed that performances follow a systematic improvement in going from semi-local to hybrid functionals.

The calculated MIR spectra need to be scaled in order to account for the limitations in the level of theory and the missing anharmonicities in the computations. In their paper (Bauschlicher et al. 2018), introduced version 3.20 of the NASA Ames PAH IR Spectroscopic Database[1] (PAHdb). In the current PAHdb database version (Boersma et al. 2014; Bauschlicher et al. 2018; Mattioda et al. 2020) the spectra are divided into three ranges: (i) C-H stretching bands ($> 2500$ cm$^{-1}$; $< 4$ μm), (ii) bands between 2500. and 1111.1 cm$^{-1}$ (i.e. between 4 and 9 μm), and (iii) bands between 1111.1 and 0. cm$^{-1}$ (i.e. $> 9$ μm); for each region a specific scale factor (0.9595, 0.9523, and 0.9563, respectively) is used to multiply the harmonic fundamentals. The computed IR spectra employing the B3LYP/4-31G model chemistry of thousands of PAHs species (including ionic species) with a number of carbons ($N_c < 400$) have been included in the PAHdb.

On the computational side, not only the prediction of vibrational spectra is subjected to multiplicative scaling factors to bring the computed normal modes to the experimental ones, but also the band strengths have to be scaled accordingly. As an example, Bauschlicher & Langhoff (1997) reported a factor of 2 difference to experimental values in the computed B3LYP/4-31G band intensities pertaining to the CH stretching modes in neutral PAHs. Other investigations, like the one by Yang et al. (2017), have scaled the B3LYP intensities of the 3.3 μm feature of some methylated PAHs to more sophisticated MP2/6-311+G(3df,3pd) ones. Given the fact that each computational method yields specific band intensities, and that investigation of astronomical PAH band strengths ratios is mainly insensitive to a selected computational method (e.g. Peeters et al. 2017), in this paper we only focus on the scaling of the predicted band positions obtained by the different computational chemistry methodologies we use.

Apart from DFT, there are very few calculations of vibrational spectroscopy with wavefunction methods. For example optimized geometries and infrared spectra were computed (Aiga 2012) for oligoacenes such as naphthalene, anthracene, naphtacene, and pentacene, PAHs, and perylene, phenanthrene, and picene using the restricted active space self-consistent field method. As their electronic ground states have an open-shell singlet multiradical character, the calculations were based on the multiconfiguration wave functions instead. The author reported good agreement for the IR spectra with experiments, after scaling all vibrations by a 0.9 factor. These wavefunction based methods are only affordable for the small sized molecules, consequently DFT has emerged as an alternative methodology for computing the electronic structure of moderate size systems.

From the computational point of view of IR spectra investigations, several procedures can be followed, among them are the widely used static quantum chemical calculations, Molecular Dynamics (MD) simulations, temperature-dependent schemes, and Monte Carlo (MC) methods. In order to better understand spectra obtained with either gas-phase or matrix-isolation experiments, theoretical anharmonic static spectra were computed to explain combination-bands and resonances in the C-H stretching region for naphthalene, anthracene, and tetracene (Mackie et al. 2015; Lemmens et al. 2019). Other models looked at the variation of band positions with temperature to extract linear slopes and quantify empirically the impact of anharmonicities (Joblin et al. 1995). Some theoretical investigations by Mulas et al. (2006) included rotational and anharmonic band structure in their MC model of the photophysics of naphthalene and anthracene. Anharmonic IR spectra of highly vibrationally excited pyrene were computed (Chen et al. 2018) using VPT2 (Mackie et al. 2015) and the Wang–Landau method in order to account for temperature effects. More recently, Chakraborty et al. (2021) studied theoretically by means of DFT-GVPT2 and also density functional tight binding (DFTB) combined with MD simulations the effects of temperature and anharmonicity in highly excited vibrational states of pyrene. The centroid and ring-polymer molecular dynamics techniques have been employed (Calvo, Parneix & Van-Oanh 2010) to naphthalene, pyrene, and coronene. There are also methods combining Tight-binding (TB) potentials with MD calculations of the electronic energy and wave functions at any particular configuration of the system. For example, TB MD and consideration of anharmonicity of the potential energy surface were used to produce absorption spectrum for naphthalene from a Fourier Transform of the dipole autocorrelation function and were found to be in good agreement with the experimental spectrum in low-temperature rare gas matrices (Van-Oanh et al. 2012). Effects of temperature and anharmonicity were investigated simultaneously in molecular dynamics studies of coronene and circumcoronene employing a PM3 electronic potential energy surface (Chen 2019). So far, dynamical calculation of vibrational spectra with classical nuclei has been reported for various gas-phase systems (Greathouse, Cygan & Simmons 2006; Schultheis et al. 2008) using parameterized potential energy surfaces in most cases, but also, more recently, surfaces based on an explicit description of the electronic structure (Margl, Schwarz & Blochl 1994; Gaigeot, Martinez & Vuilleumier 2007; Estacio & Cabral 2008). Research about extending these theoretical methodologies to account for anhamonicities to large PAHs currently constitutes a challenge as the calculations become time intensive.

In this paper, we have investigated the MIR spectral characteristics of a series of PAHs with straight edges (apart from $C_{52}H_{18}$, $C_{606}H_{68}$, $C_{706}H_{70}$, and $C_{1498}H_{108}$) and containing an even number of carbons using semi-empirical methods, PM3 and DFTB (Spiegelman et al. 2020), classical force field (FF) parameters from Amber and Gaff (Wang et al. 2004), and DFT (only for $N_c < 450$). Relevant PAHs span sizes comprising six carbons to 1498 carbons and have symmetries ranging from $D_{6h}$ to $C_s$. Specifically we consider a pool of 21 molecules $\{N_c = 6, 10, 14, 16, 24, 52, 54, 102, 190, 294, 384, 450, 506, 600, 606, 706, 806, 846, 902, 998, 1498\}$. The IR spectra computed with B3LYP/4-31G and scaled with three factors for the set from $N_c = \{10, 14, 16, 24, 52, 54, 102, 190, 294, 384\}$ are already reported in the PAHdb and will serve to validate our more approximate methodologies. The choice of the shapes and sizes of the remainder of the PAH molecules is completely arbitrary and only serves to span a wide range of large PAHs.

Current efforts to characterize and study interstellar PAHs rely heavily on theoretically predicted infrared (IR) spectra. The aim behind testing several methodologies is to be able to compute MIR spectra for large PAH molecules by reducing the computational cost. In this paper we only focus on scaling band positions and do not consider band strengths scalings. Even though we consider lower level computational methods for describing intramolecular interactions, we endeavour to reproduce the experimental bond lengths and infrared normal modes. This raises the questions about the validity and extension of semi-empirical and FFs parameter sets in producing accurate normal modes positions. Addressing the questions related to shifting band positions is at the very heart of the present paper, where the MIR spectra of 21 PAHs are analysed at different levels of theory. Anharmonic calculations are not applicable to large species and therefore are out of the scope of the present work.

The validation of the computed spectra for the small PAHs employing more approximated methods and comparison to DFT served as a benchmark for the large PAH's IR-spectra band positions prediction. In Section 2 we discuss the different theoretical methodologies employed to describe the geometrical parameters and the infrared spectra of these systems. Section 3 shows the major results obtained with DFT, semi-empirical, and FF methods, and the procedure for fitting the scale factor functions to shift band positions for any PAH size. The resulting fitting function developed for each QC method can be applied to shift any PAH molecule MIR spectra even outside the pre-selected set that served for the fitting. Shifted infrared spectra for PAHs are presented, and astrophysical implications are discussed in Section 4. The main conclusions are gathered in Section 5.

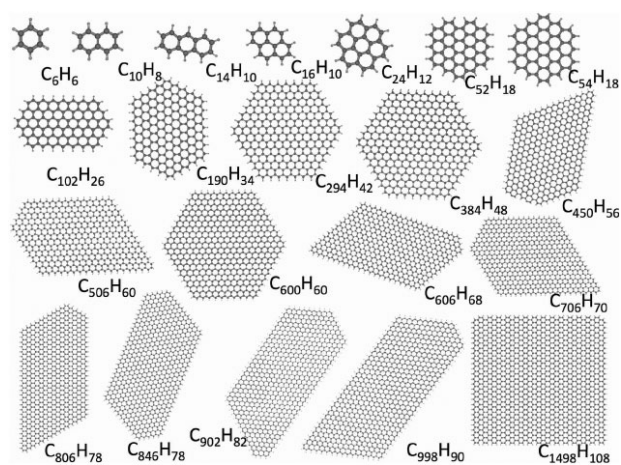

**Figure 1.** Structures of PAH molecules studied in this work. The number of carbons in the PAH molecules is given by $N_c = \{6, 10, 14, 16, 24, 52, 54, 102, 190, 294, 384, 450, 506, 600, 606, 706, 806, 846, 902, 998, 1498\}$.

## 2 METHODOLOGY

Because of computational limitations, the whole set of PAH molecules cannot be studied with the same method with either DFT or PM3. Therefore, in the present work, we employ DFT for $N_c = \{6, 10, 14, 16, 24, 52, 54, 102, 190, 294, 384, 450\}$, semi-empirical methods PM3 for $N_c = \{6, 10, 14, 16, 24, 52, 54, 102, 190, 294, 384, 450, 506, 600, 606, 706, 806\}$, DFTBA for $N_c = \{6, 10, 14, 16, 24, 450, 600, 606, 706, 806, 846, 902, 998\}$, and FF classical molecular mechanics (MM; Amber and Gaff) for $N_c = \{6, 10, 14, 16, 24, 52, 54, 102, 190, 294, 384, 450, 506, 600, 606, 706, 806, 846, 902, 998, 1498\}$ to compute infrared spectra. We perform a comparison between the spectra obtained from the different methods for the available $N_c$ values. All three types of calculations have been performed with GAUSSIAN 16 (Frisch et al. 2016) software to search for global optimizations to be sure that we have local minima, and compute harmonic vibrational spectra. Within each method, we computed the electronic energy and electric dipole moment, and their second and first derivatives respectively at the optimized geometries.

Computationally, it is challenging to extend and adapt cleverly the current methods and accurately study large molecules ($N_c > 450$) of different chemical composition, and predict their spectroscopic fingerprints. However, if this challenge is met, it will enable us for the first time to reach sizable PAH molecules and will aid in the interpretation of observed MIR spectra of a wide range of astronomical sources irrespective of their typical physical conditions (ionization, VUV, cosmic radiation, etc). For those PAHs with $N_c > 450$, whose simulation becomes exceedingly expensive when using methods like DFT due to their intrinsic cubic scaling, we evaluate the performance of more approximate methodologies, including PM3 (Stewart 1989), DFTB (Porezag et al. 1995; Seifert, Porezag & Frauenheim 1996; Elstner et al. 1998; Frauenheim et al. 2000, 2002; Oliveira et al. 2009), and the Amber and Gaff FF parameters (Wang et al. 2004), for which we adapted the charges to be able to reproduce the dipole moment accurately. Unless noted otherwise, the IR-spectra are computed for geometries that are optimized at the respective level of theory.

Fig. 1 shows the molecules studied in this work. Some of them $N_c = \{10, 14, 16, 24, 52, 54, 102, 190, 294, 384\}$ are chosen because their computed gas-phase IR spectra are available in the PAHdb database (Bauschlicher et al. 2018). Therefore, they are used as a reference to validate the accuracy of the more approximate methods.

## 2.1 DFT calculations

DFT approximations to the functionals (Hohenberg & Kohn 1964) allow the study of PAH systems with efficient computational time for molecules comprising a few hundred atoms. In the routinely used Kohn–Sham (KS) approach (Kohn & Sham 1965), the electronic problem is solved by minimizing the energy, which only depends on the electronic density, for a system of fictitious non-interacting electrons with the same electronic density and energy as the real system. Technically, this is a self-consistent mean field approach whose final energy is the same as for the fully correlated system. The structures were fully optimized and the harmonic (for $N_c$ = 6, 10, 14, 16, 24, 52, 54, 102, 190, 294, and 384) frequencies were computed using DFT.

The B3LYP hybrid density functional (Becke 1993; Stephens et al. 1994) has been extensively used by Bauschlicher et al. (2018) as it provides MIR spectra in better agreement with experiments regarding band positions than the cheaper BP86 functional.

Some publications have noted slightly different biases for the low and high harmonic frequencies (Scott & Radom 1996; Halls, Velkovski & Schlegel 2001; Yoshida et al. 2002; Sinha et al. 2004). A few have recommended scaling low and high harmonic frequencies separately (Halls et al. 2001; Sinha et al. 2004; Laury, Carlson & Wilson 2012). The B3LYP/4-31G computed and scaled harmonic frequencies by Bauschlicher & Langhoff (1997) have been reported to be in agreement with the matrix isolation experiments (Langhoff 1996) of PAH's MIR fundamental frequencies.

Pech, Joblin & Boissel (2002) and Mulas et al. (2006) found that the MIR scale factor was reasonable for the far-infrared bands as well. A far-infrared study of neutral coronene, ovalene, and dicoronylene (Mattioda et al. 2009) also showed that the MIR scale factor of 0.958 can be applied to the far-infrared spectral region as it brings the computed B3LYP/4-31G harmonic frequencies into excellent agreement with the experimental far-infrared frequencies. The same scale factor was used by Boersma et al. (2010) to bring the computed bands positions in the 15–20 μm range for 13 large molecules into the best agreement with experiment.

However, it has to be mentioned that using the B3LYP method with even basis sets, results (Yang et al. 2017) in larger computed intensities (by ~ 30 per cent) compared to experimental findings. Whereas in Bauschlicher & Langhoff (1997) it was reported that a factor 2 overestimation of the computed B3LYP/4-31G band intensities pertaining to the CH stretching modes in neutral PAH systems is to be expected.

In light of past computations by Pavlyuchko, Vasilyev & Gribov (2012) who reported that MP2/6-311G(3df,3pd) IR intensities of benzene and toluene would match experimental results, other authors (Yang et al. 2017) have scaled the computed B3LYP/6-31G(d) band intensities of methylated PAHs and their cations in the case of the 3.3 μm aromatic C−H stretch and the 3.4 μm aliphatic one by performing accurate MP2/6-311 + G(3df,3pd) calculations.

In order to correct for anharmonicities inherent to the theoretical methodologies and computational approximations, in the present B3LYP/4-31G calculations we use 'off the shelf' previously reported by Bauschlicher et al. (2018) the three harmonic scale factors to shift our B3LYP/4-31G computed spectra. However, in the case of PM3, Amber, Gaff, and DFTBA ones, we devise a fitting function to account for the scale factors for use in the MIR region.

### 2.1.1 Harmonic frequencies

As we perform static/time-independent *ab initio* methods and due to the large number of interactions/resonances in large PAHs, anharmonic IR spectra cannot be produced for big sized systems. There are essentially three different types of IR peaks: fundamental, overtone, and combination bands. In the harmonic approximation, the potential energy surface is truncated, which means that it is approximated to only second order Taylor series. It also implies approximating the dipole moment by its Taylor expansion truncated at the first order. And it is the reason why only fundamental transitions, and not combination/difference/overtone bands are obtained in this way.

Ignoring higher-order curvature terms in the potential energy function leads to deviations of the vibrational spectra predictions to those from experimental observations. Moreover, the computational combination of method and basis set choices, and also the numerical approximations may lead to severe quantitative flaw in predicting vibrational spectra.

By definition, at the harmonic level, combination bands, and overtones have zero IR intensity. Fundamental bands have inherently intense bands, make up most of the peaks seen in MIR spectra, and comprise most of the diagnostically useful peaks we have seen and will be studying for interpretation purposes. These so-called double harmonic intensities of vibrational modes are calculated using the first derivative of the dipole moment with respect to a normal coordinate, consequently, they can be method related. This kind of calculation is therefore sensitive to the accuracy of the electronic density and its variations, which in turn is the reason why DFT calculations with a basis set as small as 4-31G may yield reasonably good frequencies, modulo an empirical scaling factor, but comparatively less good intensities.

Within the harmonic approximation, molecular vibrational energy levels are evenly spaced, the peak positions in an IR spectrum are given by

$$\omega = (1/2\pi c)(K/\mu)^{1/2},$$

where $\omega$ is the wavenumber of the molecule in $cm^{-1}$, $\mathbf{K}$ the force constant, and $\mu$ the reduced mass.

After a geometrical optimization of the molecule, in the case of the harmonic approximation, the vibrational spectrum is computed from the diagonalization of the Hessian matrix.

## 2.2 Density functional TB approximations

For the large systems, starting from $N_c$ > 450 and up to $N_c$ = 1498, with B3LYP/4-31G model chemistry we were not able to perform any geometry optimizations due to computational limitations. Instead, we have employed DFTB approaches which are semi-empirical approximations to DFT. Three sets of parameters have been tested, including DFTBA (Zheng et al. 2007), MIO, and 3ob.[2] The method implemented in Gaussian, i.e. DFTBA, is derived from a second order Taylor series expansion of the density functional DFT total energy expression around a reference density. DFTBA is a version that uses analytic expressions for the matrix elements. Whereas DFTB employs the tabulated matrix elements as in the original implementation of Elstner and co-workers (Porezag et al. 1995;

---

[2]https://dftb.org/parameters/download

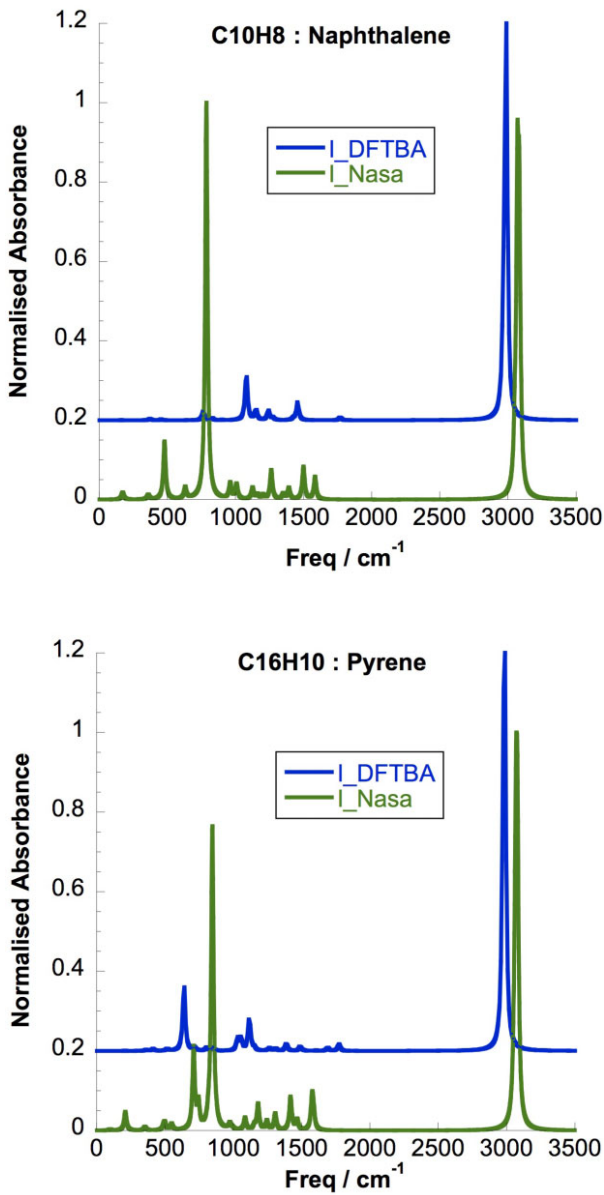

**Figure 2.** DFTBA normalized absorbance to the strongest peak vs harmonic frequencies compared to those from Ames database for naphthalene and pyrene. All spectra were convolved with a Lorentzian of FWHM equal to 20 cm$^{-1}$. DFTBA spectra were vertically shifted compared to Ames spectra for clarity.

Elstner et al. 1998). The publicly available MIO and 3ob-3-1 Slater-Koster files designed for organic molecules including respectively the MIO-1-1 (Elstner et al. 1998) and 3ob:freq-3-1 (Gaus, Goez & Elstner 2013) extensions for obtaining more accurate vibrational properties have also been used in this work.

We compare DFTBA, and the new recommended parameters (3ob:freq-3-1) for vibrational frequency studies to the MIO ones as well as to B3LYP/4-31G and the PM3 semi-empirical methods. We use DFTBA, MIO-1-1, and 3ob:freq-3-1 parameter sets to compute MIR spectra of PAHs for $N_c = \{6, 10, 14, 16, 24, 450, 600, 606, 706, 806, 846, 902, 998\}$, the bands exhibited some shifts but more importantly higher band strengths compared to DFT. Fig. 2 shows the DFTBA computed and normalized absorption spectra to the strongest peak (reported in the appendix) compared to those from Ames PAHdb in the case of naphthalene and pyrene. A similar trend for the other two sets of parameters employed with DFTB is found. The spectra were convolved with a Lorentzian of full width at half-maximum (FWHM) equal to 20 cm$^{-1}$. In Fig. 3, we also plot DFTBA absorbances normalized to the strongest peak (see Table A1) in the Appendix where we report the corresponding values) for $N_c = \{24, 384, 450, 600,$ and $606\}$ and compared to other methodologies. The discrepancies encountered in predicting DFTBA band positions might be due to the dispersion corrections that are missing in the current implementation of DFTB parameters in GAUSSIAN software. Simon and co-workers (Simon et al. 2012) obtained consistent IR anharmonic spectra when dispersion interactions are implemented in the deMonNano (Heine et al. 2009) molecular dynamics code. Also, in their paper Chakraborty et al. (2021) employ DFTB with deMonNano software for the pyrene case with a detailed assessment of its accuracy with respect to laboratory and higher level DFT calculations. They report displacement of the bands, and that their computed intensities are much more challenging to reproduce as found in our calculations.

### 2.3 Semi-empirical PM3 calculations

We have also used the PM3 semi-empirical method (Stewart 1989) for electronic structure calculations. This method was employed in the calculations of full geometry optimizations and harmonic frequency calculations. It has been shown that such methods are capable of producing reasonable vibrational spectra for large compact PAHs (Chen 2019) and high-temperature dissociation pathways for linear PAHs (Chen & Luo 2019). We note that in their MD simulations (Li 2020) have used PM3 to study the fragmentation and isomerization products of vibrationally excited coronene and its derivatives at 3000 and 4000 K, where the anharmonicity of the potential energy surface becomes important. Also, in their paper, Chen (2019) reported that anharmonic PM3 produces accurate band positions with an error less than 5 cm$^{-1}$ for the C–H stretching region in the case of naphthalene and pyrene. The author also reported measurements than those from NASA Ames PAH Database in the case of coronene and circumcoronene, where the latter shows in the case of C–H stretching modes a $\sim 20$ cm$^{-1}$ shift with respect to experiments.

In this work, we employ PM3 to circumvent the limitations related to the use of B3LYP/4-31G for large systems. From the analysis of the band intensities, we notice that for molecules up to $N_c = 54$ where all PAH structures are relatively small and have high symmetry, the PM3 computed intensities are similar to those obtained from B3LYP (see the case of $N_c = 24$ in Fig. 3). In this figure, all spectra were convolved with a Lorentzian of FWHM equal to 20 cm$^{-1}$ and we plot normalized absorbances to the strongest peak (see Table A1 in the Appendix). However, for $C_{2h}$ symmetry structures like $N_c = 102$ and 190, and also for $N_c = 294$ and 384 a stronger 3.3 μm feature, with respect to the one produced when the B3LYP is used, appears (see Fig. 3). From Fig. 4, where we plot the Mulliken point charges attributed to the PAH atoms for $N_c = 384$, we see that the charges are accumulated at the outermost benzene rings (a similar trend is noted for $N_c = 294$, not shown). In the case of ($N_c = 450, 506, 606, 706, 806$) all the bands have larger intensities (see Fig. 3), compared to smaller PAHs computed with B3LYP/4-31G. Analysis of the point charges in Fig. 4 over some of these structures ($N_c = 450$ and 606) shows that large fluctuations in the charges inside the slab appear for both PM3 and B3LYP/4-

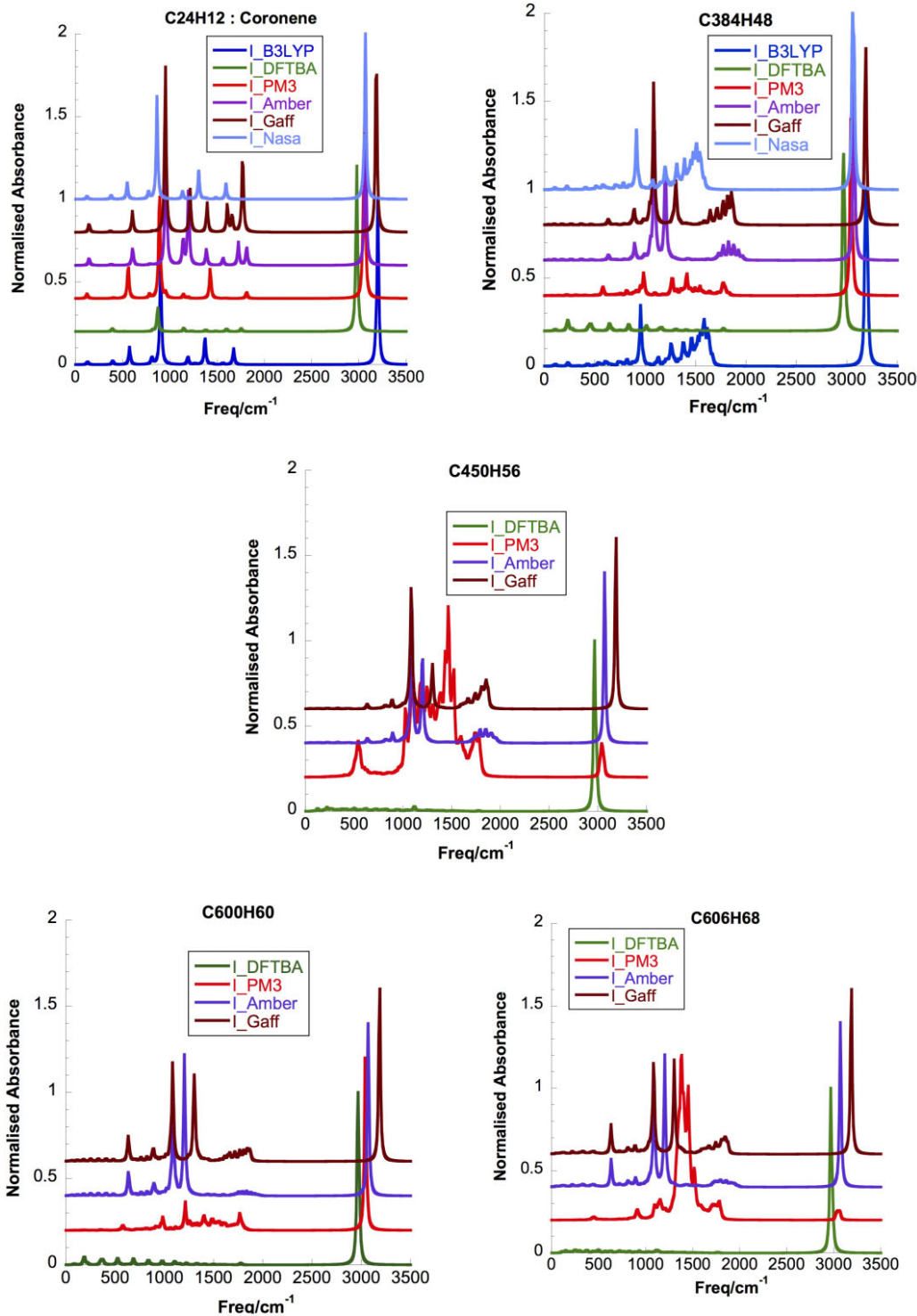

**Figure 3.** DFTBA, PM3, Amber, and Gaff computed normalized absorbances to the strongest peak *v.s.* harmonic frequencies for $N_c = 24$, 384, 450, 600, and 606 compared to available B3LYP/4-31G current results and Ames spectra when available. See similar trends in PM3 and B3LYP charges in Fig. 4, and larger C–H bond lengths obtained with PM3 compared to B3LYP in Table 1. All spectra were convolved with a Lorentzian of FWHM equal to 20 cm$^{-1}$ and have been vertically shifted for clarity.

31G computations. This will result in large C–C band intensities, and indicates that PM3 and B3LYP/4-31G Mulliken-derived charges are comparable, and that the discrepancies in the intensities of the spectra originate from the dipole moment derivation instead. From this conclusion we validate the use of PM3 computed charges in our further investigations with FF methods. For $D_{6h}$ structures ($C_{600}H_{60}$) and for $N_c = 384$ only the 3.3 $\mu$m band exhibits strong intensity as the charge fluctuations appear only around peripheral atoms, for $C_s$ structures ($C_{450}H_{56}$ & $C_{606}H_{68}$) and for $N_c \geq 384$ all bands have large intensities.

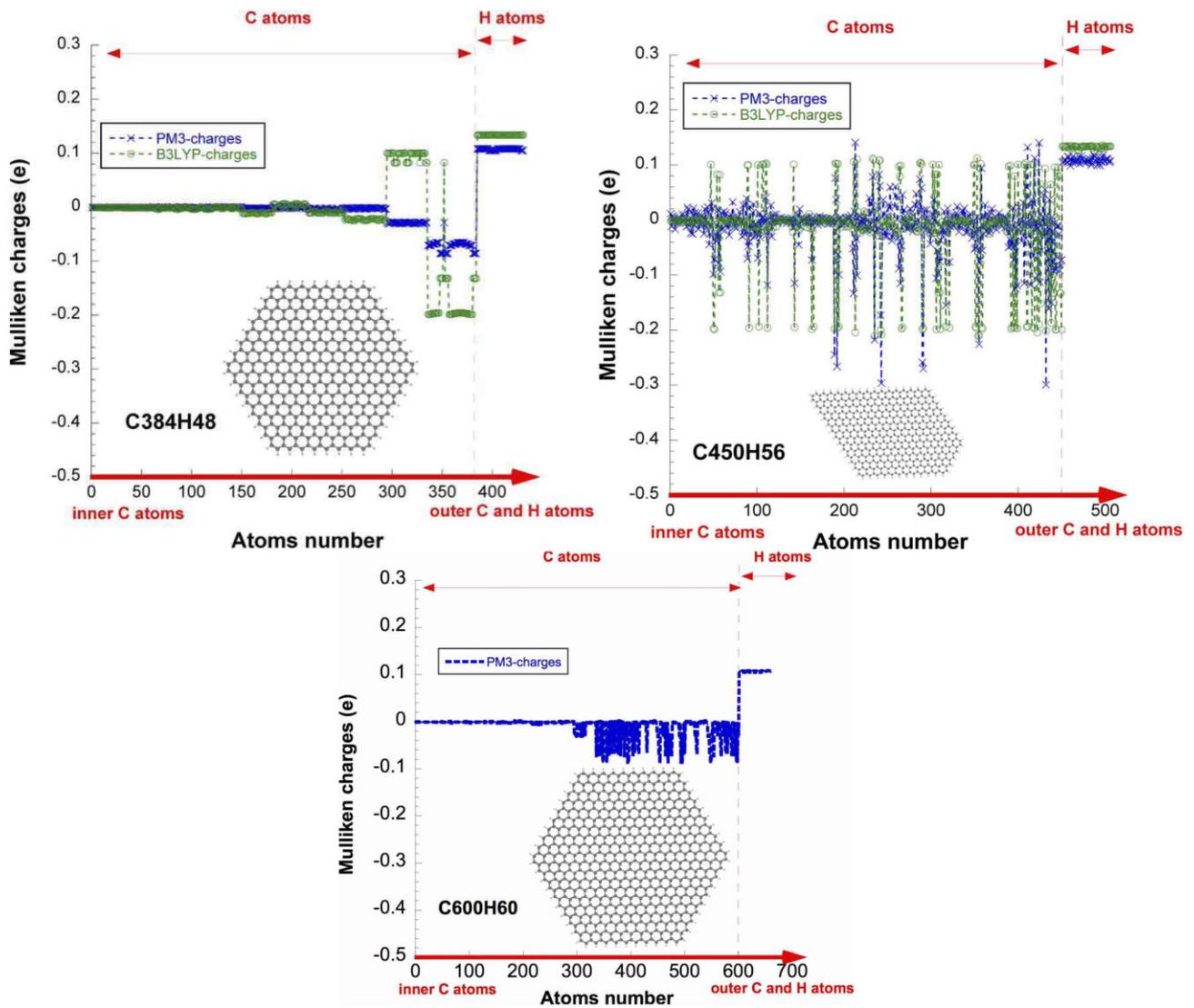

**Figure 4.** Analysis of the Mulliken point charges partitioning from B3LYP/4-31G and PM3 in the case of $N_c = 384$, 450, and 600 over the atom centres in each PAH positioned from the origin to the outer edges.

### 2.4 Amber and generalized Amber FF calculations

Vibrational and MM FF, have become predominant in computational chemistry (OPLS, Jorgensen & Tirado-Rives 1988; COMPASS, Sun 1998; Amber Case et al. 2005; CHARMM, Brooks et al. 1983). They are referred to as FFs because they use analytical potential energy functions, within the framework of classical mechanics, to predict restoring forces and the vibrational frequencies of the modelled molecules. In order to compute infrared spectra for large PAHs ($N_c > 450$), we test and employ the FF parameters in the Amber and Generalized Amber FF (Gaff; Wang et al. 2004) where Amber stands for 'Assisted Model Building with Energy Refinement' through the GAUSSIAN program (Frisch et al. 2016). The Gaff parameters (Wang et al. 2004) were designed for general organic molecules, and the atom types are much more general such that they cover most of the organic chemical space.

The expressions of the bond stretch and angle bend energy functions in Amber and Gaff are harmonic functions. All non-parabolic potential terms arise from Coulomb and van der Waals interactions. Electrostatic interactions are described via individual point charges localized at the positions of the nuclei. Amber and

Gaff provide predefined atomic charges Qeq (i.e. from charge equilibration methods; Rappe & Goddard 1991), we test them against Mulliken, and point charges from fitting to the electrostatic potential (ESP) computed with B3LYP/4-31G for smaller PAHs ($N_c \leq 450$), for larger ones PM3 Mulliken point charges are used instead. We investigate the effect of the different sets of point charges (Mulliken from B3LYP/4-31G and PM3, ESP, and Qeq) on the magnitude of the Amber computed spectra, and report this in Fig. 5. All spectra were convolved with a Lorentzian of FWHM = 20 cm⁻¹. Mulliken charges from either B3LYP or PM3 lead to similar band intensities; however, the use of ESP point charges leads to stronger features, whereas Qeq point charges result in weaker band strengths. The MM potential energy function for a given molecular system is given by (Wang et al. 2004; Case et al. 2018)

$$E_{\mathrm{MM}} = \sum_{\mathrm{bonds}} K_r (r - r_{eq})^2 + \sum_{\mathrm{angles}} K_\theta (\theta - \theta_{\mathrm{eq}})^2$$
$$+ \sum_{\mathrm{dihedrals}} \frac{V_n}{2} [1 + \cos(n\phi - \gamma)] + \sum_{i<j} \left[ \frac{A_{ij}}{R_{ij}^{12}} + \frac{B_{ij}}{R_{ij}^6} + \frac{q_i q_j}{\epsilon_{ij} R_{ij}} \right]$$

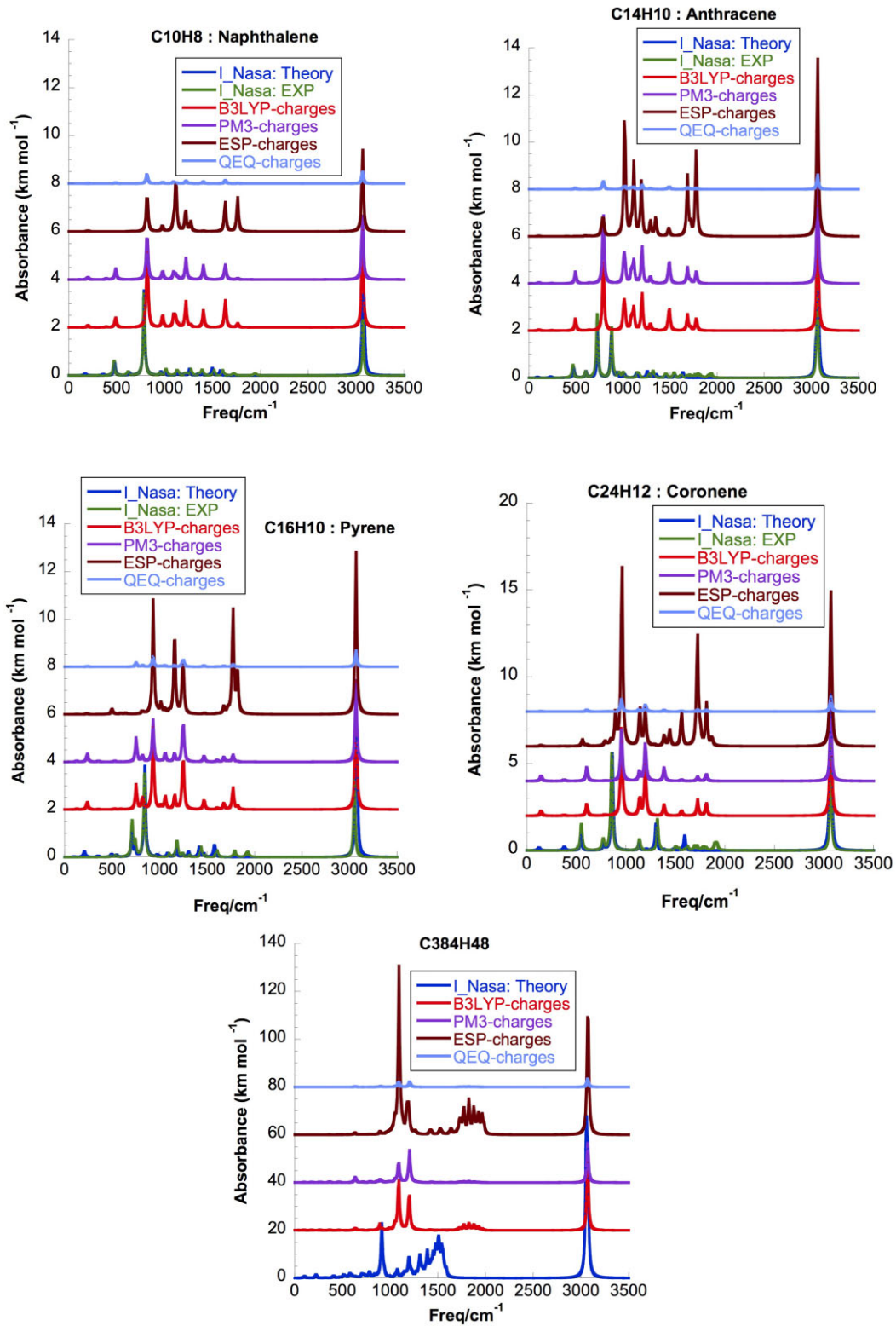

**Figure 5.** Analysis of the different charge models role on the Amber computed absorption spectra in the case of $N_c = 10, 14, 16, 24$, and 384. Comparison is made with B3LYP/4-31G and experimental (when available) spectra from NASA database. All spectra were convolved with a Lorentzian of FWHM equal to 20 cm$^{-1}$ and have been vertically shifted for clarity.

Where $K_r$ and $K_\theta$ are bond and angle force constants, respectively; $n$ is the multiplicity of the dihedral, $\gamma$ is a phase angle, $r_{eq}$ is bond equilibrium distance, $\theta_{eq}$ is angle equilibrium, and A, B, and q are non-bond parameters (Jones & Chapman 1924; Rappe et al. 1992;

Wang et al. 2004). In Gaff, equilibrium bond lengths and bond angles come from experiment and high-level *ab initio* calculations: 3000 MP2/6-31G* optimizations and 1260 MP4/6-311G(d,p) single-point calculations. The force constants are estimated through an empirical approach. Gaff uses 33 basic atom types and 22 special atom types, Carbon atoms in aromatic rings have type *ca* (sp² hybridization), while hydrogens on aromatic carbons have type *ha*.

**Table 1.** Average C−C and C−H Bonds lengths (Å) for $N_c$ = 6, 10, 14, 16, and 24 computed with B3LYP and the four basis sets $B_{x = 1, 4}$.

| $N_c$ | 6 | 10 | 14 | 16 | 24 | Average Bonds | 6 | 10 | 14 | 16 | 24 |
|---|---|---|---|---|---|---|---|---|---|---|---|
| $B_x$ | | | C−C | | | | | | C−H | | |
| 1 | 1.396 | 1.405 | 1.408 | 1.410 | 1.413 | | 1.084 | 1.084 | 1.085 | 1.084 | 1.084 |
| 2 | 1.398 | 1.407 | 1.411 | 1.412 | 1.415 | | 1.087 | 1.088 | 1.088 | 1.088 | 1.088 |
| 3 | 1.395 | 1.403 | 1.407 | 1.409 | 1.412 | | 1.084 | 1.085 | 1.085 | 1.088 | 1.085 |
| 4 | 1.394 | 1.404 | 1.407 | 1.409 | 1.412 | | 1.084 | 1.085 | 1.085 | 1.085 | 1.085 |
| Exp Herzberg 1966 | 1.397 | | | | | | 1.084 | | | | |

The validity of the FF can be tested in multiple ways. In this work it is important to compare the optimized geometrical parameters to available relatively high level *ab initio* calculations, or to those obtained experimentally to ensure the correct computation of the secondary derivatives in simulations (which are directly related to the normal modes prediction for each PAH molecule).

Although more complicated functional forms, such as polarization models, are useful for a detailed description of the molecular interactions, the power of an atomistic simulation using current force-field technology has not been completely explored yet, where a better parameterization method is one of the key issues.

Fig. 5 displays a comparison of computed spectra using different charge models to available experimental data from the literature. The performance of FF parameters with partial atomic charges from B3LYP/4-31G in test cases $N_c$ = 6-450 is encouraging in reproducing purely B3LYP/4-31G IR-spectra. In addition, the use of PM3 point charges instead of those from B3LYP/4-31G has been tested against the latter and we find similar results for band positions which is expected and exhibit systematically somehow lower intensity strengths of the bands through the whole PAHs plot (Fig. 5). For those PAHs with $N_c$ > 450, we compute Amber and Gaff spectra using partial atomic charges from PM3.

As can be seen in Fig. 3 the main noticeable difference between Amber and Gaff predicted spectra appears in the 3.3 μm band position and is related to the difference in the C−H Force constants (395.7 kcal mol⁻¹ Å⁻² for Gaff and 367.0 kcal mol⁻¹ Å⁻² for Amber).

## 3 RESULTS

### 3.1 Bond lengths at equilibrium geometry

Here we compute optimized geometries for $N_c$ = 6-1498, such as for 6 ≤ $N_c$ ≤450 we use B3LYP/4-31G, for 6 ≤ $N_c$ ≤806 we use PM3, and for 6 ≤ $N_c$ ≤998 we use DFTBA, while for $N_c$ up to 1498 only MM optimizations were computationally possible. All calculations were closed shell spin-restricted singlet states and without symmetry constraints.

In few selected cases, we report a systematic comparison of a variety of basis set combinations with B3LYP (for $N_c$ = 6, 10, 14, 16, and 24) in order to test the scale factor of the different bands. Specifically, we compared calculations from four basis sets ranging in complexity 4-31G (B1), 6-31+G(d) (B2), 6-311G(d,p) (B3), 6-311++G(d,p) (B4) as can be seen in Table 1. From the table where we report average C–C and C–H bond lengths, we see that the bond lengths with the four basis set combinations agree with each other.

Comparison of experimental bond lengths to the computed values is within 0.01 Å.

We next address the relative performances of the PM3 and FF methods for calculating the bond lengths. In Table 2 we report average C−C and C−H bond lengths computed with PM3, and Amber and we compare them to our B3LYP/4-31G data for PAH sizes in the range $N_c$ = 6-450. In any case, the difference between the B3LYP/B1, PM3, and Amber results for the C−C and C−H bonds remains below a 2 per cent. Gaff computed bond lengths show a similar trend to those obtained from Amber. We also compute the mean and RMS differences of the C−C and H−H bond lengths for each species compared to DFT for Amber and PM3 methods as can be seen in Fig. 6. We give the mean difference of means and RMS of RMS's over the set of species to get unbiased statistics towards the larger molecules, this is reported in Table 3. In any case, the means and RMS's differences do not exceed 2 per cent and remain constant across the $N_c$ ensemble.

### 3.2 Matching the harmonic frequencies

When we employ the three scale factors mentioned in Bauschlicher et al. (2018), and apply them to our B3LYP/4-31G harmonic fundamentals we reproduce exactly those from PAHdb for $N_c$ = {10, 14, 16, 24, 52, 54, 190, 197, 294, and 384}; this can be seen from the red ratio between the Amesdb scaled vibrational modes $\nu_i$ and our computed ones $\nu_{i, Ames}/\nu_{i, B3LYP}$ that remains constant as a function of frequency in Fig. 7.

Here we focus on finding specific scaling factors for each QC method ∈ {PM3, DFTBA, Amber, Gaff}. For this purpose we started off by applying the NASA factors (Bauschlicher et al. 2018). This has been proven unsuccessful. Therefore, we have considered instead ratios between the NASA scaled normal modes and those from a given QC method. We notice that when there are only relatively few modes (not a large molecule), it is possible to do some matching by eye, visualizing the modes, but this becomes an impossible task when there are many similar modes. Therefore, we tried to implement an automated technique to match the normal modes by vector and not by value. For most cases it is possible to match the DFTBA and PM3 frequencies and the vector matching is relatively good and puts modes with similar frequencies together. However, matching modes by the vectors resulted in complete failure for Amber and Gaff. For example, in the Amber data for benzene, one of the modes of the degenerate pair at 609 cm⁻¹ has a very high (> 90 per cent) match with a non-degenerate B3LYP mode at 1026 cm⁻¹, which is clearly unphysical. Several attempts to ensure that the degenerate pairs were matched in a proper way (i.e. accounting for the underlying symmetry) were also unsuccessful and we concluded that the vector data were not of high enough quality for the matching to work properly. To conclude, we found that the best way forward was to sort the modes by vector

**Table 2.** Average C−C and C−H Bonds lengths (Å) for selected $N_c$ values obtained with B3LYP/4 − 31G, PM3, and Amber.

| $N_c$ | 6 | | 10 | | 14 | | 16 | | 24 | |
|---|---|---|---|---|---|---|---|---|---|---|
| | C−C | C−H | C−C | C−H | C−C | C−H | C−C | C−H | C−C | C−H |
| B3LYP/B1 | 1.396 | 1.084 | 1.405 | 1.084 | 1.408 | 1.085 | 1.410 | 1.084 | 1.413 | 1.084 |
| PM3 | 1.391 | 1.095 | 1.399 | 1.096 | 1.403 | 1.096 | 1.406 | 1.096 | 1.409 | 1.096 |
| Amber | 1.406 | 1.081 | 1.408 | 1.080 | 1.408 | 1.080 | 1.408 | 1.080 | 1.409 | 1.080 |

| $N_c$ | 52 | | 102 | | 190 | | 294 | | 384 | |
|---|---|---|---|---|---|---|---|---|---|---|
| | C−C | C−H | C−C | C−H | C−C | C−H | C−C | C−H | C−C | C−H |
| B3LYP/B1 | 1.417 | 1.084 | 1.419 | 1.085 | 1.420 | 1.085 | 1.421 | 1.085 | 1.421 | 1.085 |
| PM3 | 1.413 | 1.097 | 1.415 | 1.097 | 1.417 | 1.097 | 1.418 | 1.097 | 1.418 | 1.097 |
| Amber | 1.409 | 1.079 | 1.409 | 1.079 | 1.410 | 1.079 | 1.410 | 1.079 | 1.410 | 1.079 |

| $N_c$ | 450 | | 506 | | 606 | | 706 | | 806 | |
|---|---|---|---|---|---|---|---|---|---|---|
| | C−C | C−H | C−C | C−H | C−C | C−H | C−C | C−H | C−C | C−H |
| B3LYP/B1 | 1.421 | 1.085 | – | – | – | – | – | – | – | – |
| PM3 | 1.418 | 1.097 | 1.418 | 1.097 | 1.418 | 1.097 | 1.418 | 1.097 | 1.418 | 1.097 |
| Amber | 1.410 | 1.079 | 1.411 | 1.08 | 1.411 | 1.08 | 1.411 | 1.08 | 1.411 | 1.08 |

| $N_c$ | 846 | | 998 | | 1498 | |
|---|---|---|---|---|---|---|
| | C−C | C−H | C−C | C−H | C−C | C−H |
| B3LYP/B1 | – | – | – | – | – | – |
| PM3 | 1.418 | 1.097 | 1.418 | 1.097 | 1.419 | 1.099 |
| Amber | 1.411 | 1.08 | 1.411 | 1.08 | 1.411 | 1.078 |

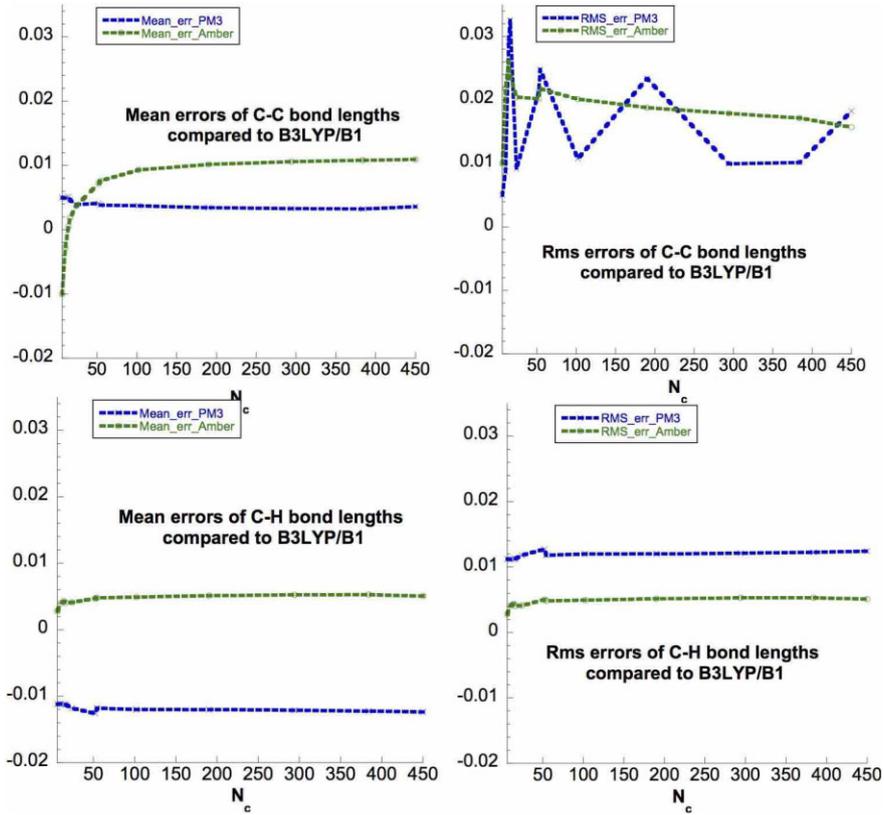

**Figure 6.** RMS and mean errors of the C−C and C−H bond lengths compared to B3LYP/B1 for PM3 and Amber as a function of $N_c$.

for DFTBA and PM3 while for Amber and Gaff we sorted the modes by frequency.

Fig. 7 displays ratios for Amber, Gaff, PM3, and DFTBA; they show noisier behaviour based on the sole three-scale factors in the case of Ames database DFT spectra. We also can see denser data points for $N_c = 384$ because of the higher number of normal modes compared to $N_c = 16$. A smooth numerical fit can easily be applied; in this work we employed instead a polynomial of degree three to give an analytical function to the fit. The degree 3 polynomials are also plotted to provide a fit to the ratios.

**Table 3.** Mean difference of means and root-mean-square (RMS) of RMS's over the set of species $N_c$ = 6 up to 450 for PM3, and Amber compared to B3LYP/B1.

|       | *Mean difference of means* | | RMS of RMS's | |
|-------|---------|---------|---------|---------|
|       | C−C     | C−H     | C−C     | C−H     |
| PM3   | 0.004   | -0.012  | 0.019   | 0.012   |
| Amber | 0.005   | 0.005   | 0.019   | 0.005   |

## 3.3 Scale factor derivation for PM3, DFTBA, Amber, and Gaff computed PAHs spectra

In this section, we investigate for each QC method employed to compute the harmonic infrared spectra i.e. PM3, DFTBA, Amber, and Gaff the necessary shifts that have to be applied for the bands in the 300–3500 cm$^{-1}$ range. For our computed B3LYP/B1 MIR spectra, and $N_c$ = {6,10,14,16,24,52,102,190,294,384} we apply the three NASA scale factors mentioned previously (Bauschlicher et al. 2018). For a given $N_c$, we have (3 × ($N_c$ + $N_H$)-6) normal modes of vibrations $v_{i,\,calc}$, and for each method we plot the ratio between the Amesdb scaled vibrational modes $v_i$ and our computed ones $v_{i,\,calc}$. This ratio can be seen in Fig. 7 for $N_c$ = 16, and 384, and derived for the four abovementioned methods. One can see from the figure that this ratio presents fluctuating behaviour versus $v_i$ except for B3LYP/B1 data; the observed trend is expected for the semi-empirical and empirical methods employed here as their associated general errors are larger than for the *ab initio* counterparts because they are designed largely to give energies for the class of compound for which they are parameterized. The second point that emerges from the same figure is that for PM3, DFTBA, Amber, and Gaff, the ratio remains identical through the whole $N_c$ range {6, 10, 14, 16, 24, 52, 102, 190, 294, 384} i.e the same ratio for a given QC method for all $N_c$ values. This finding has allowed us to conduct a polynomial fitting for each of the ratios, i.e. ratio $\frac{v_{i,\,Ames}}{v_{i,\,calc}}$ versus $v_i$, and lead to the development of a degree three polynomial $\mathcal{P}_3 = \mathcal{M}_0 + \mathcal{M}_1 \times v_i + \mathcal{M}_2 \times v_i^2 + \mathcal{M}_3 \times v_i^3$ capable of shifting the calculated normal modes $v_{i,\,calc}$. We report one polynomial for each of the PM3, DFTBA, Amber, and Gaff computed IR spectra which can also be seen in the same figure with dotted lines.

We next investigate the $N_c$ dependence of the $\mathcal{P}_3$ coefficients i.e. $\mathcal{M}_{j=0-3}$. For this purpose, we plot in Fig. 8 $\mathcal{M}_{j=0-3}$ versus $N_c$ for Amber (similar trends were obtained for PM3, DFTBA, Gaff polynomial coefficients), where we see that all $\mathcal{M}_j$ values reach convergence vs $N_c$ starting from 50 to 150, which means that the $\mathcal{M}_j$ coefficients become constant for large $N_c$ and take the value of the Morse function at the asymptotic region. The figure clearly shows that they can be extrapolated to very large $N_c$ values. We provide an analytical fitting to a Morse function of all $\mathcal{M}_{j=0-3}$ relative to all four methods, the resulting $\mathcal{FSF}$ of the two variables $N_c$ and $v_i$ can be given by the following equation:

$$\mathcal{FSF}(N_c, v_i) = \mathcal{M}_0(N_c) + \mathcal{M}_1(N_c) \times v_i + \mathcal{M}_2(N_c) \times v_i^2 + \mathcal{M}_3(N_c) \times v_i^3, \quad (1)$$

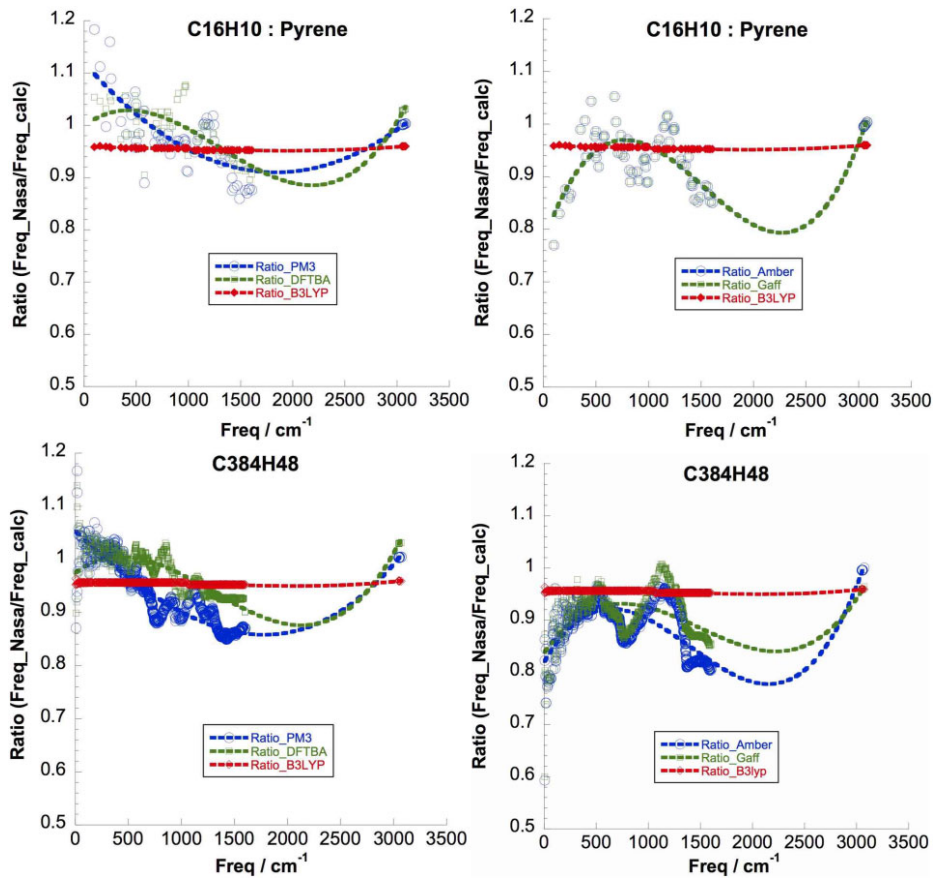

**Figure 7.** Ratio between the NASA B3LYP scaled frequencies and those from the present B3LYP/B1, PM3, DFTBA, Amber, and Gaff given by symbols vs harmonic frequencies with their corresponding $\mathcal{P}_3$ given by dashed lines for $N_c$ = 16 and 384.

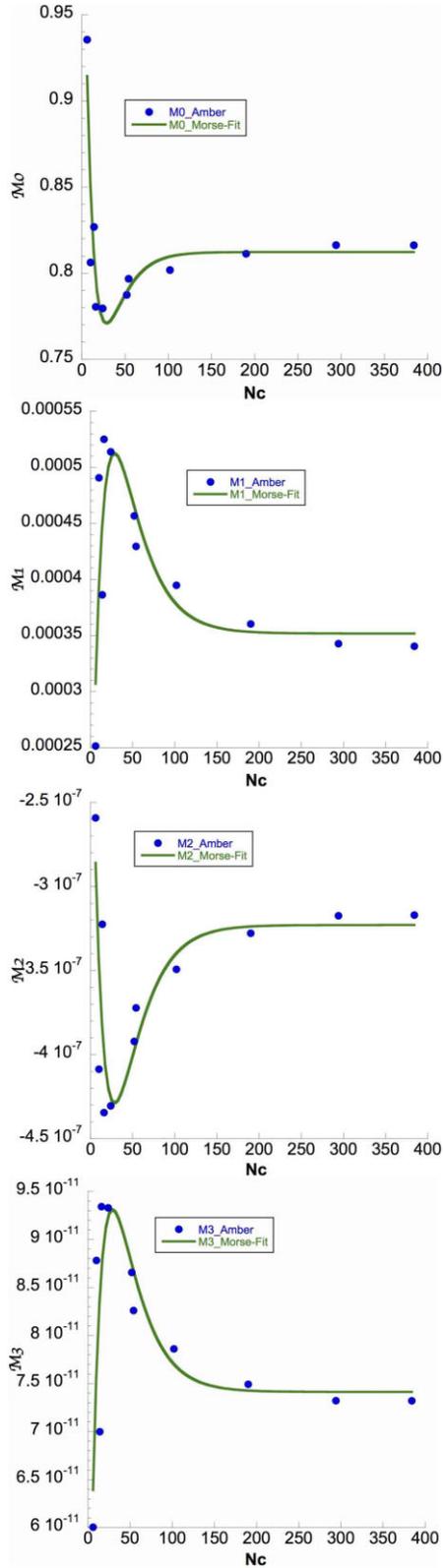

**Figure 8.** $M_i$ ($i = 0$–$3$) versus $N_c$ and the corresponding Morse function fits in the case of Amber data.

where

$$\mathcal{M}_j(N_c) = a_{1,j} \times [1 - \exp(-a_{2,j} \times (N_c - a_{3,j}))]^2 + a_{4,j}, \quad j = 0\text{ - }3. \tag{2}$$

The $a_{\ell,j}$ parameters for $\ell = 1 - 4$, and $j = 0 - 3$, are given in Table 4, while Table 5 gives the RMS and variance of residuals of all $\mathcal{M}_{j=0-3}$. fitted parameters.

Our fitting procedure provides parameters applicable to PM3, DFTBA, Amber, and Gaff calculated frequencies. The two-variable functions we have developed work very well for all the methods and one can use any of them to predict scaled band positions for any employed QC method and any given PAH size provided the appropriate parameters in Table 4 are used. As has been detailed in Section 2.3 the B3LYP/B1 and PM3 charges are very similar therefore with Amber and Gaff, we employ B3LYP/B1 Mulliken charges for ($N_c \leq 450$), for larger PAHs PM3 Mulliken point charges are used instead. From Fig. 8 we see that the matching between the fit and the computed data points is improved for $N_c \geq \sim 50$, and becomes constant for $N_c$ values larger than 200, irrespective of whether we use PM3, DFTBA, Amber, or Gaff. The parameters of the scaling function for each QC method depicted in Table 4 were derived from only the 10 PAH molecules $N_c = \{6, 10, 14, 16, 24, 52, 102, 190, 294, 384\}$ that are in the PAHdb database, where $N_c = 54$ was not included. To validate the use of this function for a given $N_c$ not included in the pre-selected set, we applied it to shift the MIR spectra for $N_c = 54$ (for which NASA data are available) and we display its corresponding polynomial parameters $\mathcal{M}_{0,1,2,3}$ in Fig. 8 where we clearly see that they fall on the Morse fit in the case of Amber calculations. This has also been verified for the other methodologies. The fitting function can also be applied to $N_c > 384$ as it has been noticed that for $N_c > 200$ the polynomial parameters become constant. This comes with no surprise as for large PAHs the surface area becomes extended with respect to the PAH edges and the geometry resembles that of infinite carbon sheets. For this reason, the electronic density distribution becomes more uniform when the PAH size increases, and as a consequence the dipole moment too, resulting in constant shifts as has been applied to PAH molecules with $N_c > 384$. This methodology is therefore extendable to predict shifted MIR spectra for large PAHs where DFT becomes exceedingly time intensive. For $N_c$ values $\leq 200$, sophisticated combinations of DFT methodologies with large basis sets can be easily applied to compute harmonic and also anharmonic frequencies.

## 4 DISCUSSION

We report the calculated spectra with PM3, DFTBA, Amber, and Gaff after applying the $\mathcal{FSF}$ to shift the MIR bands. Furthermore, the spectra were convolved with a Lorentzian of FWHM $= 20\,\mathrm{cm}^{-1}$. We calculate the maximum attained temperature from the provided energy photon input (i.e. 12 eV) as detailed in Boersma et al. (2014) and subsequently use Kirchoff's law to estimate the emission spectrum from the absorption one, by multiplying a blackbody at that fixed temperature with the integrated cross-section of each vibrational transition.

In Fig. 9 we plot the current emission spectra results (convolved and exposed to a 12 eV radiation field) from PM3, DFTBA, Amber, and Gaff calculations and compare them with Ames data when available in the case of $N_c = 24, 52, 384$, showing how the efficiency of emitting into different emission features depends on the PAH size. We note an acceptable agreement for some PM3 features this has already been reported in Section 2.3 where we have shown that the discrepancies in the intensities of the spectra originate from the dipole moment derivation (large charge fluctuations inside the less symmetric PAH molecules appear for both PM3 and B3LYP/B1 computations). Furthermore, the matching is less good in the case of FFs, and the intensities are very discrepant for DFTBA probably also due to the missing dispersion corrections in the current implementation of this methodology in the software we use.

**Table 4.** Fitting parameters for the Frequency Scaling Function $\mathcal{FSF}(N_c, \nu_i)$ derived for computed PM3, Amber, Gaff, and DFTBA harmonic frequencies.

| $\mathcal{P}_3$ parameters: | $\mathcal{M}_0$ | $\mathcal{M}_1$ | $\mathcal{M}_2$ | $\mathcal{M}_3$ |
|---|---|---|---|---|
| Morse parameters | | | | |
| | | PM3 | | |
| $a_1$ | $3.21782 \times 10^{-3}$ | $-5.51485 \times 10^{-5}$ | $3.47968 \times 10^{-8}$ | $-5.2452 \times 10^{-12}$ |
| $a_2$ | $4.3827 \times 10^{-2}$ | $3.03191 \times 10^{-2}$ | $2.90544 \times 10^{-2}$ | $3.0222 \times 10^{-2}$ |
| $a_3$ | $52.8202$ | $39.4997$ | $40.1874$ | $41.7789$ |
| $a_4$ | $1.04954$ | $-1.07174 \times 10^{-4}$ | $-2.29886 \times 10^{-8}$ | $1.7277 \times 10^{-11}$ |
| | | Amber | | |
| $a_1$ | $4.14598 \times 10^{-2}$ | $-1.60775 \times 10^{-4}$ | $1.0597 \times 10^{-7}$ | $-1.89854 \times 10^{-11}$ |
| $a_2$ | $4.66733 \times 10^{-2}$ | $3.3637 \times 10^{-2}$ | $3.4011 \times 10^{-2}$ | $3.48681 \times 10^{-2}$ |
| $a_3$ | $28.5174$ | $28.49$ | $28.663$ | $29.132$ |
| $a_4$ | $7.70913 \times 10^{-1}$ | $5.12631 \times 10^{-4}$ | $-4.28844 \times 10^{-7}$ | $9.31214 \times 10^{-11}$ |
| | | Gaff | | |
| $a_1$ | $3.59961 \times 10^{-2}$ | $-1.58737 \times 10^{-4}$ | $1.26796 \times 10^{-7}$ | $-2.61409 \times 10^{-11}$ |
| $a_2$ | $6.06731 \times 10^{-2}$ | $3.40613 \times 10^{-2}$ | $2.91348 \times 10^{-2}$ | $2.75969 \times 10^{-2}$ |
| $a_3$ | $23.5934$ | $24.8307$ | $21.3158$ | $18.2988$ |
| $a_4$ | $7.97694 \times 10^{-1}$ | $4.39382 \times 10^{-4}$ | $-3.60125 \times 10^{-7}$ | $7.56412 \times 10^{-11}$ |
| | | DFTBA | | |
| $a_1$ | $2.0848 \times 10^{-2}$ | $-1.44977 \times 10^{-4}$ | $1.111 \times 10^{-7}$ | $-2.35309 \times 10^{-11}$ |
| $a_2$ | $2.1559 \times 10^{-2}$ | $9.61885 \times 10^{-2}$ | $2.31499 \times 10^{-2}$ | $4.68213 \times 10^{-2}$ |
| $a_3$ | $37.455$ | $13.6896$ | $17.0286$ | $18.798$ |
| $a_4$ | $9.948 \times 10 - 1$ | $1.88071 \times 10^{-4}$ | $-2.19466 \times 10^{-7}$ | $5.86621 \times 10^{-11}$ |

**Table 5.** RMS and variance of residuals (reduced $\chi^2$) for PM3, Amber, Gaff, and DFTBA corresponding to the fitted $\mathcal{M}_{j=0-3}$ parameters.

| $\mathcal{P}_3$ parameters: | $\mathcal{M}_0$ | $\mathcal{M}_1$ | $\mathcal{M}_2$ | $\mathcal{M}_3$ |
|---|---|---|---|---|
| | | PM3 | | |
| RMS | $6.46 \times 10^{-3}$ | $1.07 \times 10^{-5}$ | $7.06 \times 10^{-9}$ | $1.46 \times 10^{-12}$ |
| $\chi^2$ | $4.17 \times 10^{-5}$ | $1.15 \times 10^{-10}$ | $4.98 \times 10^{-17}$ | $2.12 \times 10^{-24}$ |
| | | Amber | | |
| RMS | $6.91 \times 10^{-3}$ | $1.81 \times 10^{-5}$ | $1.22 \times 10^{-8}$ | $2.37 \times 10^{-12}$ |
| $\chi^2$ | $4.78 \times 10^{-5}$ | $3.29 \times 10^{-10}$ | $1.49 \times 10^{-16}$ | $5.59 \times 10^{-24}$ |
| | | Gaff | | |
| RMS | $7.29 \times 10^{-3}$ | $1.97 \times 10^{-5}$ | $1.40 \times 10^{-8}$ | $2.93 \times 10^{-12}$ |
| $\chi^2$ | $5.32 \times 10^{-5}$ | $3.90 \times 10^{-10}$ | $1.96 \times 10^{-16}$ | $8.61 \times 10^{-24}$ |
| | | DFTBA | | |
| RMS | $4.73 \times 10^{-3}$ | $2.26 \times 10^{-5}$ | $5.53 \times 10^{-9}$ | $2.84 \times 10^{-12}$ |
| $\chi^2$ | $2.23 \times 10^{-5}$ | $5.14 \times 10^{-10}$ | $3.06 \times 10^{-17}$ | $8.07 \times 10^{-24}$ |

We introduce in Fig. 10 a sample of normalized absorption spectra for different molecular sizes $\{N_c = 24, 52, 384, 806, 1498\}$; in the same plot, we show the same sample of normalized emission spectra exposed to 12 eV radiation field. All spectra are computed with the same Amber method, convolved with a Lorentzian of FWHM equal to 20 cm$^{-1}$, and shifted with the $\mathcal{FSF}$ Amber parameters given in Table 4. Such a comparison allows us to disentangle the effect of size ($N_c$) from that of the heat capacity on the strength of PAH bands and, subsequently, on PAH band ratios. As can be seen from the plots, the strength of the 3.3 PAH band decreases with increasing $N_c$ both in the absorption spectra as well as in the emission spectra. However, the 11.3 and the 17.0 PAH bands increase with increasing $N_c$ and this trend stays the same in the absorption and emission spectra as well. From this comparison it is clear that PAH band ratios such as the 3.3/11.3 and/or the 3.3/17.0 can be used to infer the size of molecules responsible for the observed PAH emission as we discuss later on.

In Fig. 11 we compare the current Amber emission spectra for $N_c = 1498$ to $N_c = 52$ that are exposed to two radiation fields of 5 and 12 eV, respectively. One can see that the intensity of the different emission bands varies depending on the energy of the photon that was absorbed. However, the ratio of the intensities of the different features would be less sensitive to changes in the photon energy as shown in Rigopoulou et al. (2021). Comparing these spectra at the two photon energies highlights the specific effect of including the infrared spectrum of the large molecule in the model. The adapted shifting parameters give an acceptable match in the band positions. Probably a better matching could be achieved by applying a smooth spline fit or an interpolation instead of polynomials. Nevertheless, the developed $\mathcal{FSF}$ bands shifting tool provides a general trend of all PAH band positions in the MIR regime. The same protocol can be adapted and used to investigate the sensitivity of the MIR band positions for astrophysically important data.

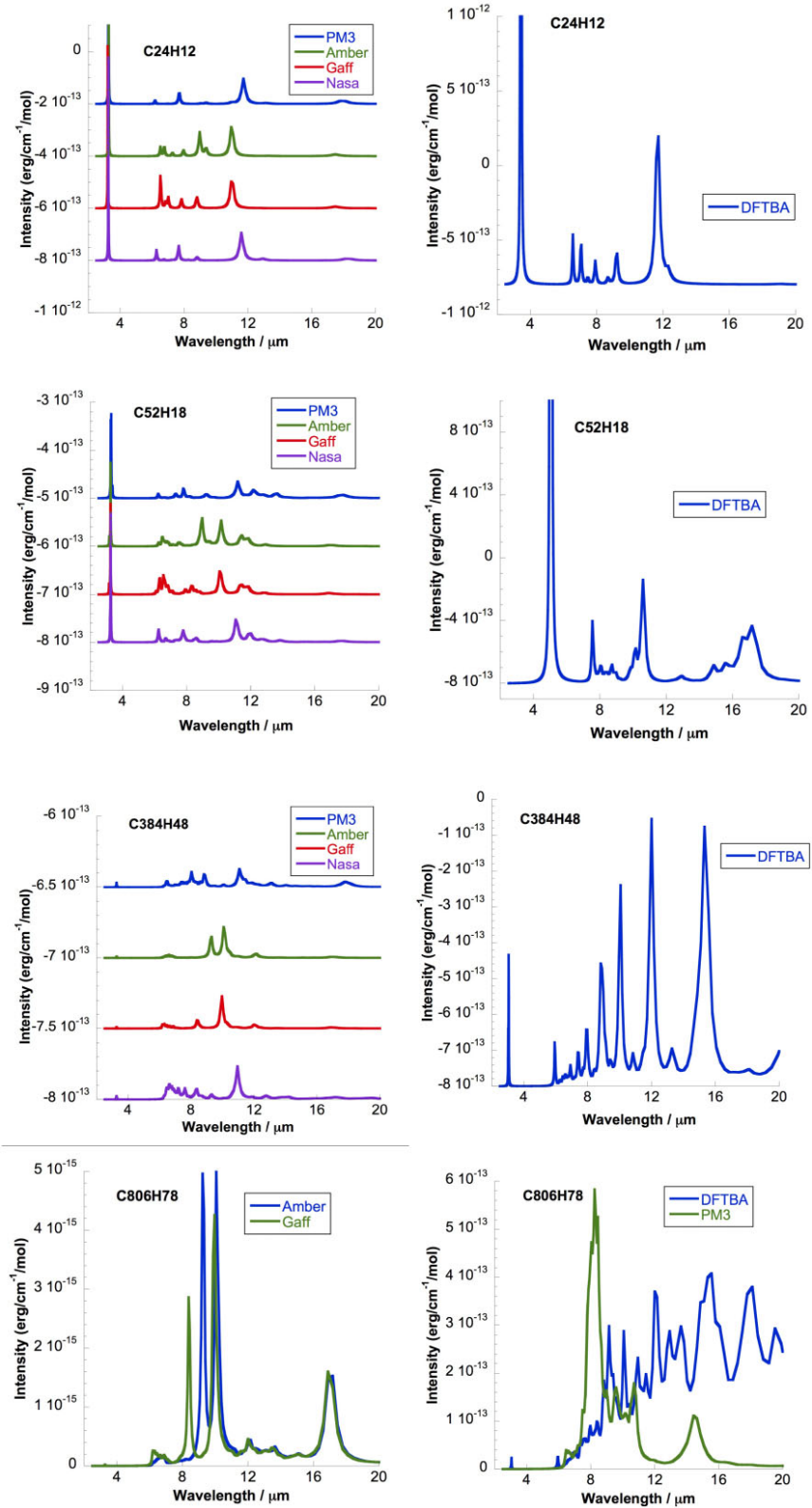

**Figure 9.** $\mathcal{FSF}$-shifted, exposed to a 12 eV radiation field, and convolved emission spectra for $N_c$ = 24, 52, 384, and 806 computed with PM3, DFTBA, Amber, and Gaff, compared to B3LYP/B1 shifted spectra from NASA database for $N_c$ = 24, 52, and 384. The spectra have been vertically shifted for clarity.

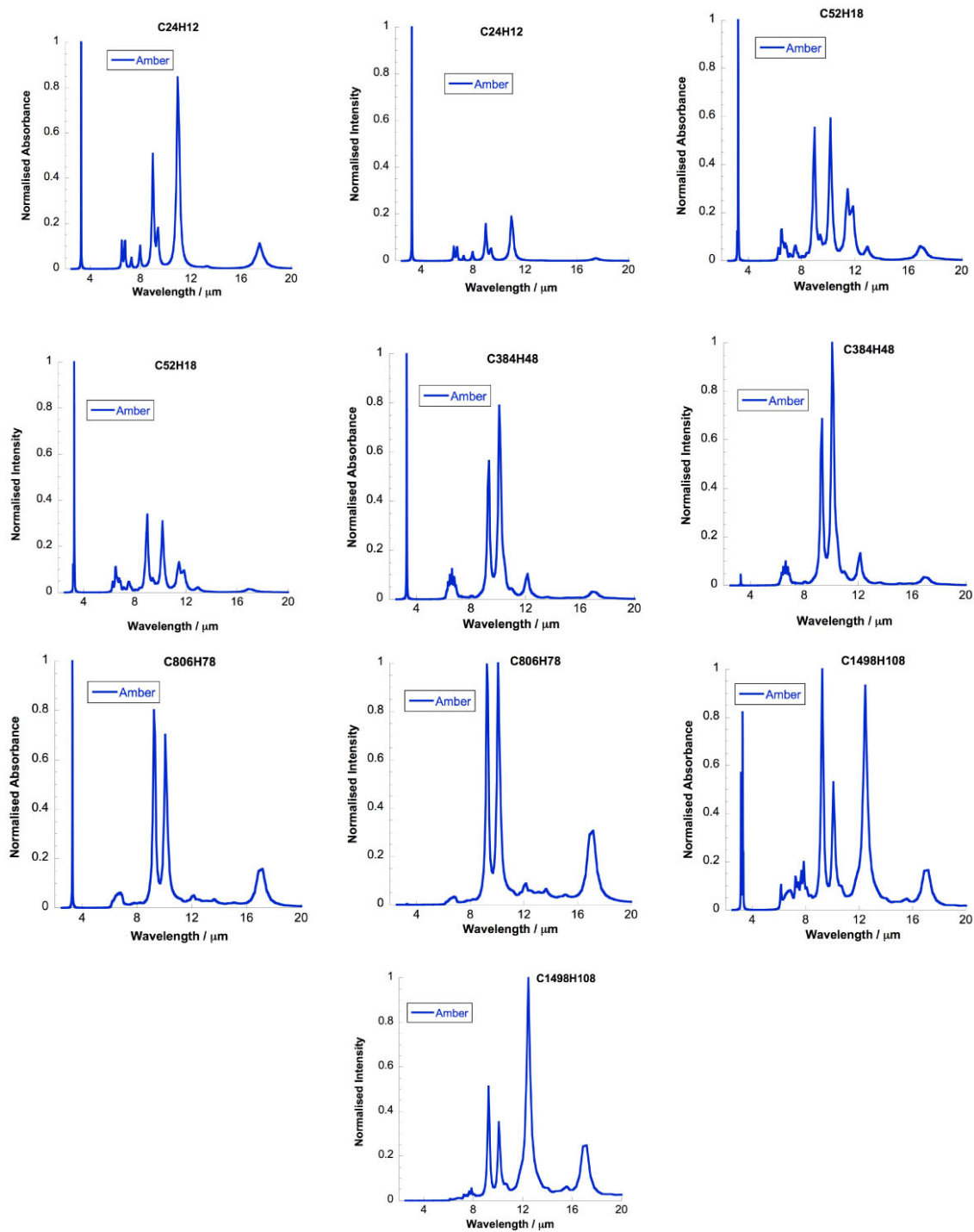

**Figure 10.** Normalized emission spectra exposed to a 12 eV radiation field compared to normalized absorption spectra for $N_c = 24, 52, 384, 806$, and 1498. All spectra were computed with Amber, convolved with a Lorentzian of FWHM equal to 20 cm$^{-1}$, shifted with $\mathcal{FSF}$, and normalized to the strongest peak.

This work has enabled us to extend predictions of MIR spectra positions for PAH molecules with $N_c > 384$, i.e beyond what is currently available in the literature. The proposed strategy provides a correct comparison as for the general trend even though we do not succeed to provide a perfect agreement in the sub-features frequencies as well as intensities. These investigations show that even though the features inside the spectra change with respect to $N_c$ values, the trends of the MIR band positions is preserved. In particular, the 3.3 μm band weakens for large PAHs, while the 11.3 and 17.0 μm ones become prominent. The known PAH MIR emission bands span two opposite ends from the 3.3 μm band to the 17μm complex. They allow hence to trace PAH size distribution, and the variation between the two bands therefore allows us to probe the overall size shift in different interstellar environments as demonstrated in (Lai et al. 2020).

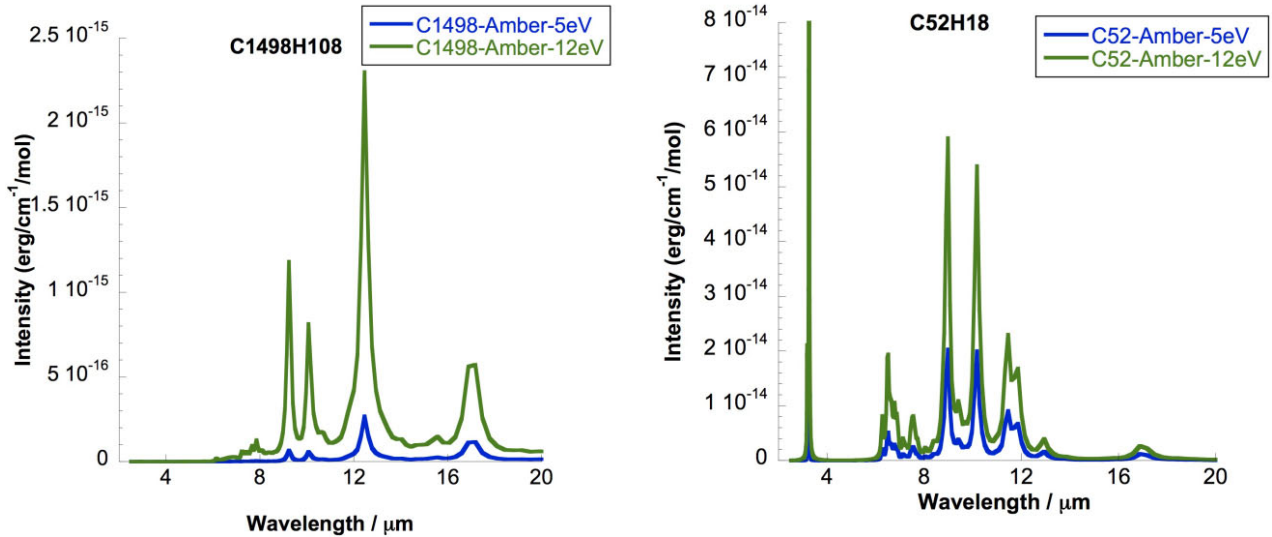

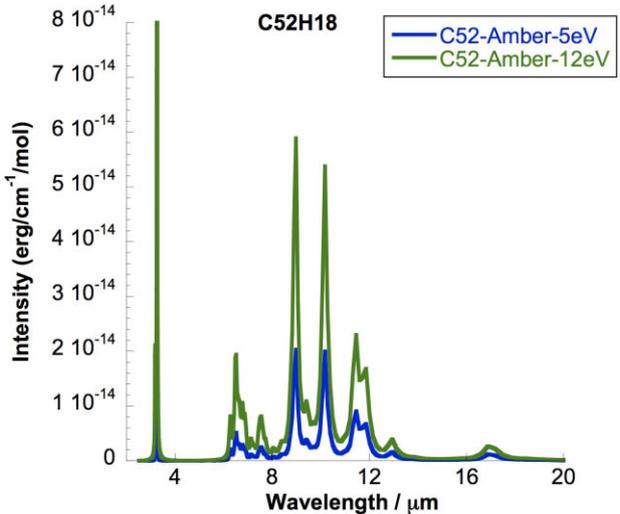

**Figure 11.** $\mathcal{F}S\mathcal{F}$-shifted, and convolved emission spectra for $N_c = 1498$ and 52 computed with Amber, the spectra were exposed to two radiation fields 5 eV (blue line) and 12 eV (green line). One can see the weakening of the 3.3 μm feature in $N_c = 1498$, and the appearance instead of the 17.0 μm band.

It is well established (e.g. Schutte, Tielens & Allamandola 1993; Draine & Li 2007) that the 3.3 μm PAH feature is attributed to the smallest PAH molecules (those with $N_c < 100$) whereas PAH molecules with $N_c > 1000$ are mainly responsible for the emission of the bands that contribute towards the 17.0 μm complex PAH. Therefore, the ratio of these two bands should be a useful indicator of the overall size of PAHs from 100 to 1000 in a variety of astrophysical environments. Indeed, Lai et al. (2020) presented a comparison of the relative changes of the 3.3 and 17.0 μm bands for a sample of normal star-forming galaxies. While they note similar trends in the fractional decrease of the 3.3 μm feature with increasing luminosity of their sample galaxies, this trend is further complicated by the fact that in astrophysical environments the 3.3 μm PAH feature is susceptible to extinction as well as metallicity. The combination of 17.0 and 3.3 μm PAHs may also be able to shed light into a possible fragmentation process among the PAH population such as top down fragmentation theories where large molecules are broken into smaller PAHs when exposed to intense radiation fields (as suggested by e.g. Jones 2016). Such a process will result in a redistribution of PAH sizes.

On the other hand, earlier investigations of PAHs in low metallicity objects (such as dwarf galaxies, e.g. Madden et al. 2006; Gordon et al. 2008; Hunt et al. 2010) have suggested that only large PAHs are able to survive the harsh environments. But the detection of strong 3.3 μm emission in IIZw40 (Lai et al. 2020) may imply that the opposite is true, namely the survival of small grains. Such a scenario would favour a bottom-up formation process.

Our work extending the available vibrational PAH spectra into molecules containing $N_c \sim 1500$ is crucial for the following reasons: on the one hand we can now investigate the trends in the 3.3/17.0 μm PAH ratio as a function of PAH size (a ratio which has been less well studied so far compared to the more widely used 6.2/7.7 and/or 3.3/11.3), but also and perhaps more importantly we can build a detailed model of PAH size distribution and how this responds to exposure to fields of different radiation intensities.

In Fig. 12 we show spectra taken with the Short Wavelength Spectrometer (SWS) on board the Infrared Space Observatory (*ISO*) covering the main mid-infrared PAH bands. We chose two sources that sample different interstellar radiation fields: at the low excitation end we show the SWS01 spectrum of the reflection nebula NGC 2023 (Verstraete et al. 2001). At the high excitation end we show the SWS01 spectrum of the Orion Bar (Verstraete et al. 2001) a well-studied high excitation source. The spectra were taken from the compilation of Sloan et al. (2003). The physical conditions prevailing in each of the two sources are reflected in the strength of the PAH features and the ensuing PAH band ratios. In the case of the low excitation source the 3.3 (11.3) PAH feature is weaker (stronger) in comparison to those seen in the Orion Bar. Likewise, we note that the 17 μm PAH band is more prominent in large PAH molecules compared to smaller ones. This is similar to the situation in Fig. 12 where we show the theoretical spectra of small ($C_{52}H_{18}$) and large ($C_{1498}H_{108}$) PAH molecules which are exposed to two different radiation fields of 5 and 12 eV. The spectrum of NGC 2023 is qualitatively similar to the theoretical spectra exposed to a 5 eV radiation field, contrary to what is seen in the spectrum of the Orion Bar. The prediction of the new large PAH spectra we computed in this paper will therefore enable us to better constrain the properties of the underlying sources in upcoming *James Webb Space Telescope* (*JWST*) spectroscopic observations.

Our computational efforts rely partly on the forthcoming improved spatial resolution and increased sensitivity observations (by more than an order of magnitude) with *JWST* to better understand these issues about the multirange specific shifts and assignment of the bands. It is evident that accurate modelling of the process that leads to excitation and cooling of the PAH molecules and subsequent prediction of the resulting emission spectra is an imperative task in order to interpret PAH spectra. The recent launch of *JWST* offers a unique opportunity to detect PAHs from a wide range of objects with unprecedented resolution. The present theoretical quantum chemical calculations can provide MIR spectral information for a wide range of systems. An important and largely unexplored topic in the astrochemical context is a theoretical framework to describe the physical properties of large-scale PAH macromolecules which is a timely question given what is inferred from the analysis of astronomical observations. We provide at the following Oxford University Link[3] all the IR spectra computed either with PM3, Amber, or Gaff with their XYZ ASCII files.

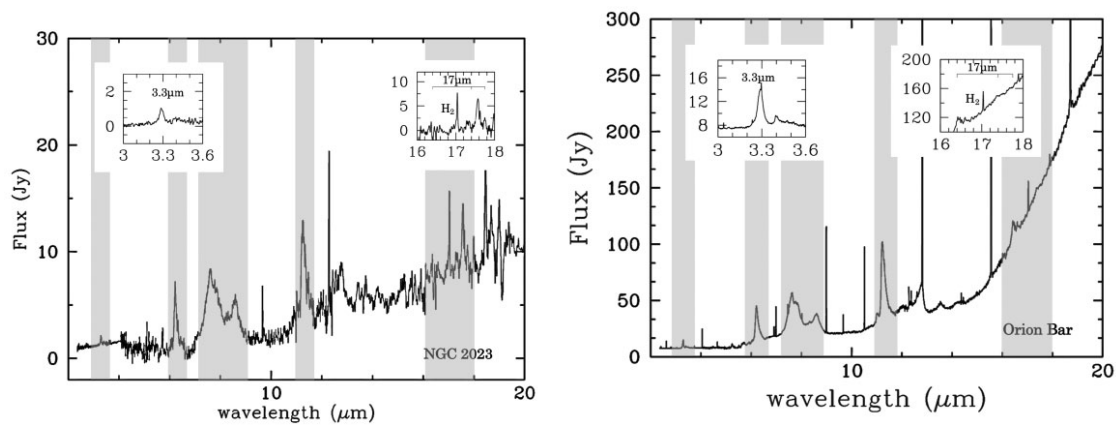

**Figure 12.** *ISO*-SWS01 PAH spectra of the reflection nebula NGC 2023 (left-hand panel) and of the Orion Bar (right-hand panel). The shaded regions indicate the location of the 3.3, 6.2, 7.7, 8.6, 11.3, and the 17 μm PAH bands. For clarity, the inset panels zoom in on the 3.3 and the 17 μm PAH bands. The narrow unresolved lines correspond to various fine structure and $H_2$ lines (spectra from Verstraete et al. 2001).

## 5 CONCLUSIONS

We report in this paper optimized geometries and vibrational frequency analysis for a range of PAH molecules and compare them to literature/experiments when available. Our computed B3LYP/4-31G IR spectra employ three scale factors for PAHs with $N_c < 450$ as dictated in the literature.

We focused on extending the existing computational methodologies to predict large PAHs MIR spectra by using semi-empirical and empirical tools and developed a function form which depends on frequencies $v$ and number of carbons $N_c$ to account for band shifts to match the B3LYP/B1 Ames database results. In summary, a new theoretical approach to account for shifts of vibrational spectra computed with PM3, DFTBA, Amber, and Gaff for large sized polycyclic aromatic hydrocarbons has been reported in this paper. We employ the Amber and Gaff parameters with B3LYP (for $N_c \leq 450$) and PM3 (for $N_c > 450$) computed Mulliken point charges rather than the Qeq ones. This tool has allowed a detailed study of the normal-mode frequencies in the neutral PAHs. The method is validated for PAHs up to $N_c = 384$ for which shifted B3LYP/4-31G results exist, and was extended to predict MIR spectra for up to $N_c = 1498$. The development of a two-variables function to account for band shifts has produced shifted B3LYP/4-31G shifted band positions retrieved from the Ames database. The function has therefore served to predict accurate general trends for all MIR bands observed in astronomical sources for PAH sizes up to $N_c = 1498$ for the first time. The trend predicted in the large PAH molecules set shows that the 3.3 μm weakens while the 11.3 and 17.0 μm become more distinguishable with their stronger peak magnitude. The data we make available via the url link[3] will be very useful for the wider community and can be employed in astrophysical models of the emission from specific objects in space. It is worth mentioning that the current predicted PM3 spectra exhibit complex band intensities for some large $N_c$ values, while DFTB calculations resulted in large fluxes compared to the other QC methods and some shifted features particularly the 3.3 μm band that we find shifted to higher wavelengths. An optimized parameterization of the FF potential for aromatic systems has been tested, but did not lead to any improvement over the existing FF parameters and consequently was not pursued any further.

## ACKNOWLEDGEMENTS

BK is thankful to the John Fell Fund, University of Oxford and to the AfOx Visiting Fellowship. The calculations have partly been performed at the University of Oxford Advanced Research Computing (ARC) HPC service. The work has also benefited from the Project HPC-EUROPA3 (INFRAIA-2016-1-730897), with the support of the EC Research Innovation Action under the H2020 Programme; in particular, BK gratefully acknowledges the support of the Astrophysics Department at Oxford University and the computer resources and technical support provided by the Irish Centre for High-End Computing (ICHEC). DR acknowledges support from the John Fell Fund, University of Oxford and the Science and Technology Facilities Council (STFC) through grant ST/S000488/1.

## DATA AVAILABILITY

This publication makes use of data products from Ames NASA PAHdb database at https://www.astrochemistry.org/pahdb/theoretical/3.20/default/view. The data underlying this article which consist in ASCII files for all PM3, Amber, Gaff, and DFTBA spectra for all studied PAH molecules are available in [the Oxford University Astrophysics sub-Department Repository], at https://users.ox.ac.uk/~phys2144/PAH-IR-DATA/

## APPENDIX A: INTENSITIES OF THE STRONGEST BANDS THAT APPEAR IN THE NORMALIZED PLOTS SHOWN IN THE PAPER

**Table A1.** List of the strongest peak position (Freq) in $(cm^{-1})$ and its absorbance value (Abs) in $(km\,mol^{-1})$ used for the normalization of convolved spectra (shown throughout the paper) with a Lorentzian of FWHM of 20 $cm^{-1}$.

| Method | DFT | | DFTBA | | PM3 | | Amber | | Gaff | |
|---|---|---|---|---|---|---|---|---|---|---|
| | Freq | Abs | Freq | Abs | Freq | Abs | Freq | Abs | Freq | Abs |
| $C_{10}H_8$ | | | 2987.7 | 105. | | | | | | |
| Ames | 787. | 3.53 | – | – | – | – | – | – | – | - |
| $C_{16}H_{10}$ | | | 2987.7 | 370. | | | | | | |
| Ames | 3073 | 5.01 | – | – | – | – | – | – | – | - |
| $C_{24}H_{12}$ | | | 2977.7 | 402. | 3057.9 | 7.27 | 3067.9 | 6.6 | 953.14 | 5.67 |
| Ames | 3198.2 | 8.56 | – | – | – | – | – | – | – | - |
| $C_{384}H_{48}$ | | | 2967.7 | 16900 | 3037.8 | 90.6 | 3067.9 | 28.6 | 3188.2 | 28.6 |
| Ames | 3188.2 | 64.3 | – | – | – | – | – | – | – | - |
| $C_{450}H_{56}$ | | | 2967.7 | 32100 | 11464.3 | 987. | 3067.9 | 33.1 | 3188.2 | 33.4 |
| Ames | 3057. | 67.7 | – | – | – | – | – | – | – | - |
| $C_{600}H_{60}$ | | | 2967.7 | 34200 | 3037.8 | 202. | 3067.9 | 20.5 | 3188.2 | 20.1 |
| $C_{606}H_{68}$ | | | 2967.7 | 47400 | 1384.1 | 2310.0 | 3067.9 | 23.9 | 3188.2 | 23.4 |

This paper has been typeset from a TEX/LATEX file prepared by the author.